\documentclass[manuscript]{aastex}

\slugcomment{Accepted for publication in The Astronomical Journal}

\shorttitle{Taurus Galaxy}
\shortauthors{Clemens et al.}

\begin{document}

\title{NEAR-INFRARED POLARIMETRY OF A NORMAL SPIRAL
GALAXY VIEWED THROUGH THE TAURUS MOLECULAR CLOUD COMPLEX}

\author{Dan P. Clemens, M. D. Pavel\altaffilmark{1}, and L. R. Cashman}
\affil{Institute for Astrophysical Research, Boston University,
    725 Commonwealth Ave, Boston, MA 02215 USA}
\altaffiltext{1}{Current address: Department of
Astronomy, The University of Texas, 1 University Station, C1400, Austin, TX 
78712-0259, USA}
\email{clemens@bu.edu, pavelmi@utexas.edu, lcashman@bu.edu}

\begin{abstract}
Few normal galaxies have been probed using near-infrared polarimetry, even though
it reveals magnetic fields in the cool interstellar medium better than either optical
or radio polarimetry. 
Deep $H$~band (1.6~$\mu$m)
linear imaging polarimetry toward Taurus serendipitously included the galaxy 
2MASX~J04412715+2433110 with adequate sensitivity and 
resolution to map polarization across nearly its full extent. 
The observations revealed the galaxy
to be a steeply inclined ($\sim 75$\degr) disk type with a diameter, encompassing
90\% of the Petrosian flux, of 4.2 kpc at a distance of 53~Mpc. 
Because the sight line passes through the Taurus Molecular 
Cloud complex, the foreground polarization needed to be measured 
and removed. 
The foreground extinction A$_{\rm V}$ of 2.00$\pm$0.10~mag  and 
reddening $E(H - K)$ of 0.125$\pm$0.009~mag were also assessed and removed, 
based on analysis of 2MASS, 
UKIDSS, {\it Spitzer}, and {\it WISE} photometry
using the NICE, NICER, and RJCE methods. 
Corrected for the polarized
foreground, the galaxy polarization values range
from zero to 3\%. 
The polarizations are dominated by a disk-parallel magnetic field geometry, especially to the northeast, while either a vertical field or
single scattering of bulge light produces disk-normal polarizations
to the southwest. 
The multi-kpc coherence of the magnetic field revealed by the infrared 
polarimetry is in close agreement
with short wavelength radio synchrotron observations of edge-on 
galaxies, indicating that both cool and warm interstellar media of disk
galaxies may be threaded by common magnetic fields.

\end{abstract}


\section{Introduction}

What roles do magnetic fields play in the interstellar media (ISM) of galaxies, and
how are those fields created and sustained? How common are galaxy-wide magnetic 
fields in normal galaxies like the Milky Way? Radio observations have revealed a wealth
of information about magnetic fields by tracing total and polarized synchrotron 
emission and Faraday
rotation in the warm, diffuse gas in galaxies \citep[see reviews by][]{Beck96,Beck09}, but there 
are few observational techniques capable of probing magnetic fields in the cooler, star-forming 
ISM in normal galaxies \citep[e.g.,][]{Jones97, Packham11}.

The radio observations generally show disk-parallel (toroidal) magnetic fields for 
edge-on disk galaxies \citep[e.g.,][]{Sukumar91,Dumke95,Krause06,Soida11},
with transitions to more vertical (poloidal) fields in the halos. Optical wavelength
polarizations, originally thought to arise from dichroic (selective) extinction 
\citep{Serkowski75} that
traced magnetic fields \citep[e.g.,][]{Scarrott96}, show polarization orientations 
perpendicular to galaxy disks. \citet{Fendt96} carefully compared their optical polarimetry
to radio synchrotron polarimetry for three highly-inclined disk galaxies and concluded
that single scattering dominates at optical wavelengths. As a result, magnetic
information is not revealed by optical polarimetry of such galaxies. In the near-infrared (NIR),
scattering is less important \citep{Wood97a,Wood97b}, enabling NIR polarimetry to
reveal magnetic fields in the cool, star-forming ISM that radio synchrotron emission
and polarization often appear to avoid \citep{Beck09}.

The Galactic Plane Infrared Polarization Survey \citep[GPIPS:][]{Clemens12a} is in the 
process of using background NIR starlight polarization at $H$ band (1.6~$\mu$m) to probe the 
magnetic field of the star-forming mid-plane
of the northern Milky Way. GPIPS reveals the field to distances in the disk of about 5 -- 7 kpc,
on arcmin scales. To put GPIPS findings in context 
and to enable testing models of the Milky Way magnetic field 
\citep{Pavel11,Jansson12a,Jansson12b}, NIR polarimetric
observations of external, Milky Way analog galaxies are needed. 
Only a few edge-on systems have been previously polarimetrically detected \citep{Jones89} or partially mapped \citep{Jones97} in the NIR. Recent observations of M51, a face-on
grand-design spiral previously mapped with optical polarimetry \citep{Scarrott87}, failed to detect 
$H$ band polarization, to quite low levels \citep{Pavel12}.  
Infrared polarimetry of active galaxies, such as M82 \citep{Bingham76, Jones00} and
others \citep{Ruiz00, Ruiz03}, have revealed complicated, 
multi-origin polarization components, but the degree to which these objects inform 
our understanding of the magnetic fields in more normal galaxies like the Milky Way is unclear.

A new small area, deep $H$ band polarization survey of a portion of the Taurus 
Molecular Cloud complex (distinct from the study of \citealt{Chapman11})
was undertaken 
to examine relationships between magnetic fields and turbulence \citep{Falgarone13}.
Serendipitously, one resolved disk galaxy was apparent in the observations.
The galaxy proved bright enough to measure near-infrared (NIR) polarization
across almost its full extent. 
Given how very few galaxy observations in the NIR meet those conditions, 
a more detailed analysis was warranted.

However, because this galaxy (2MASX~J04412715+243311, designated hereafter
as the `T-Galaxy' for its location in Taurus), is viewed through the gas,
dust, and magnetic fields associated with the Taurus Molecular Cloud complex
\citep{Heyer87}, it was necessary to calibrate and remove the foreground Galactic effects 
on the polarization and photometric observations. This was challenging because the 
expected T-Galaxy $H$ band polarization and the foreground Taurus
polarization were likely to have similar percentage values. Described in the following
sections are the new $H$ band polarization observations, the data
reduction, the general properties determined for the T-Galaxy from the 
$H$ band photometry, the determination of the Taurus region foreground 
polarization and extinction, the correction to the T-Galaxy polarization 
observations, and analysis of the nature of the magnetic field present in the 
T-Galaxy and how its field relates to those in other normal galaxies.

\section{Observations and Data Reduction}

\subsection{Polarimetry}

Ground-based near-infrared $H$ band (1.6~$\mu$m) imaging polarimetry observations 
were conducted on the night of 2011 October 13 (UT) using the Mimir instrument
\citep{Clemens07} on 
the 1.83 m Perkins telescope, located outside Flagstaff, AZ.  The observations consisted
of four sets of 96 images, each of 10~s exposure time. Each set consisted of six sky dither
positions, distributed about a hex-pattern of diameter 15 arcsec, with sixteen images at each
dither position. Those sixteen images were obtained through sixteen distinct rotation angles of
the cold, compound half-wave-plate (HWP) to modulate the incoming linear polarization
stellar signals and were analyzed by a cold, fixed wire grid prior to detection by the
$1024 \times 1024$ pixel InSb Aladdin III array. They provided a four-fold redundancy for calculating
the four independent polarization position angles (IPPAs: 0, 45, 90, 135\degr) used to
form the total intensity normalized linear polarization Stokes $U$ and $Q$ 
percentages\footnote[2]{Stokes 
values in this paper are formally $Q/I$ and $U/I$, but are designated $Q$ and
$U$ to reflect conventional usage at optical and infrared wavelengths.}. 
The pixel size was 0.58 arcsec 
and the full image size was $10 \times 10$ arcmin. The details of the data collection methodology
follow closely those employed for GPIPS, as described in \citet{Clemens12a}.

\begin{figure}
\includegraphics[ trim = 1.2cm 1cm 0cm 1.0cm, clip=true, scale=0.80]{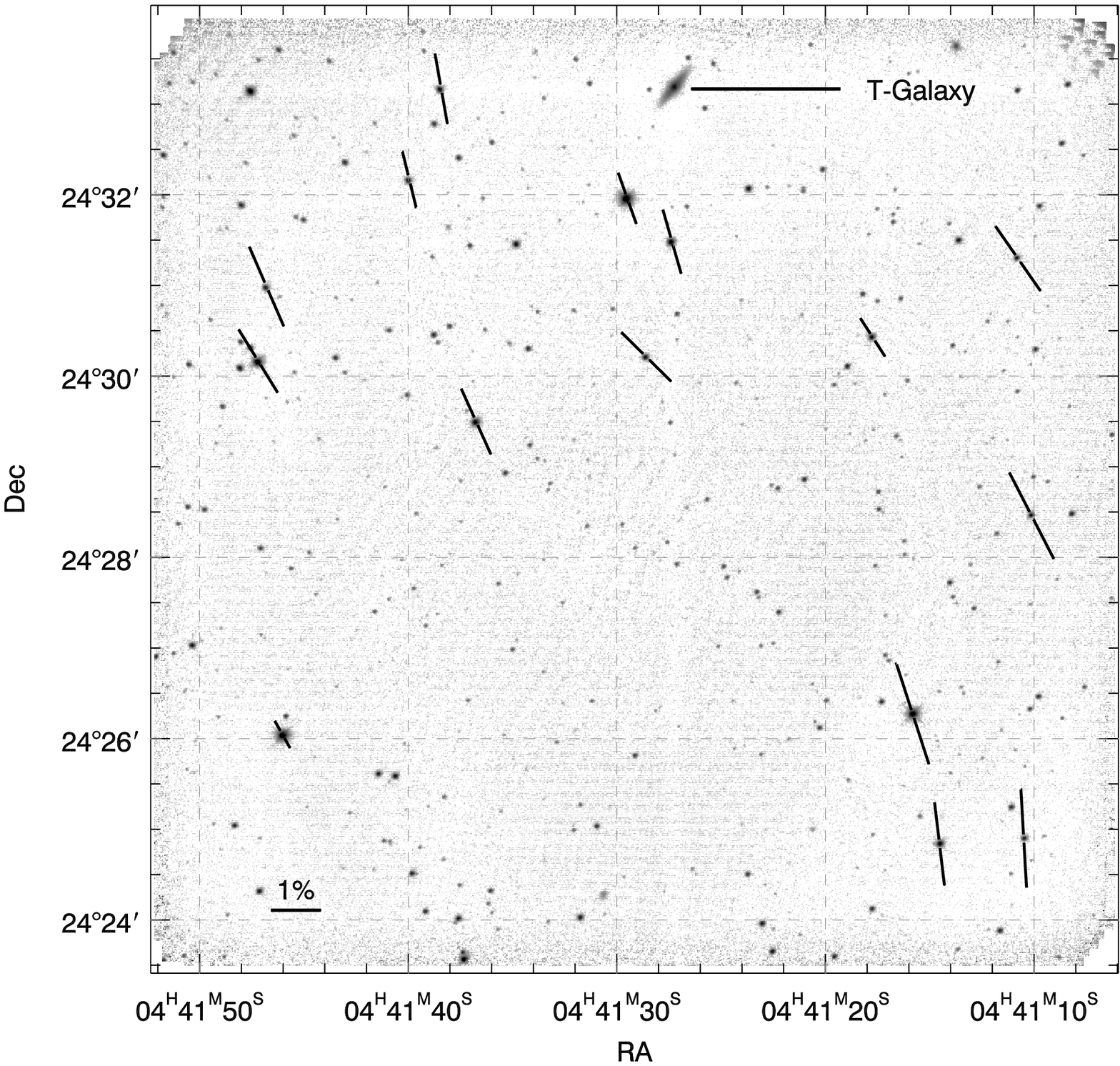}
\caption{Combined Mimir $H$ band Stokes $I$ image toward the field in Taurus containing the T-Galaxy.
Linear polarizations are shown as black vectors for fifteen stars in the field exhibiting
low polarimetric uncertainties. Vector lengths represent percentage polarization $P$, with
a reference length shown in the lower left. Vector angles represent the polarization position
angle, P.A., measured East from North in this equatorial representation in J2000 coordinates.
\label{fig_deep_image}}
\end{figure}

Reduction of the Mimir data also followed the GPIPS pipeline approach \citep{Clemens12a},
which included detector non-linearity correction, HWP angle-specific flat-fielding, 
and dark current corrections 
as well as corrections for sky transmission variations and instrumental polarization. The images
were background-corrected in two ways. For analysis of the stars in the field,
development of `super-sky' corrections were used to model and remove
the background from the images, as is
normally done for GPIPS data. For analysis of the T-Galaxy, the same images
had background zones free of emission identified and those pixels were used
in a second order fit to model and remove the background near the 
T-Galaxy. This avoids inadvertent removal of faint, extended emission from the T-Galaxy
that super-sky use might cause,
given the small dither offsets used. All of the images 
were combined to yield a Stokes $I$ (total intensity) image, shown in  
Figure~\ref{fig_deep_image}. The T-Galaxy appears in the top middle of the Figure.
The stars have FWHM values of $3.00 \pm 0.04$~pixels, or 1.75 arcsec.

GPIPS stellar polarization values are normally calculated from point spread function (PSF)
cleaned aperture photometry 
of the stars found in the 16 HWP images for each field, after registering and averaging
the corresponding HWP images from all dithered observations. 
Stars in the T-Galaxy field were analyzed for their polarization properties using
the normal GPIPS stellar processing, which yielded a catalog of polarizations, positions, and 
$JHK$ photometry for the stars that matched to 2MASS \citep{Skrutskie06} ones. 
A total of 239 stars had 
polarization measurements completed, though less than 10\% of those were significantly detected 
in polarization.
This low $P$ detection rate is due to the low numbers of stellar photons counted
for the fainter stars and the background levels present in the $H$ band. 
Figure~\ref{fig_deep_image} shows the polarization
properties of fifteen field stars exhibiting low polarization uncertainties 
($\sigma_{\rm PA} < 15 \degr$, but also filtered such that $\sigma_{\rm P} 
< 1$~\%). The 200+ polarization non-detection stars are still useful,
as their Stokes $U$ and $Q$ percentages are gaussian distributed, and may
be averaged, with weighting, to yield significant polarimetric information, as described
in Section 4.1, below.

\begin{figure}
\includegraphics[ trim = 1.25cm 1cm 0cm 1.5cm, clip=true, scale=0.95]{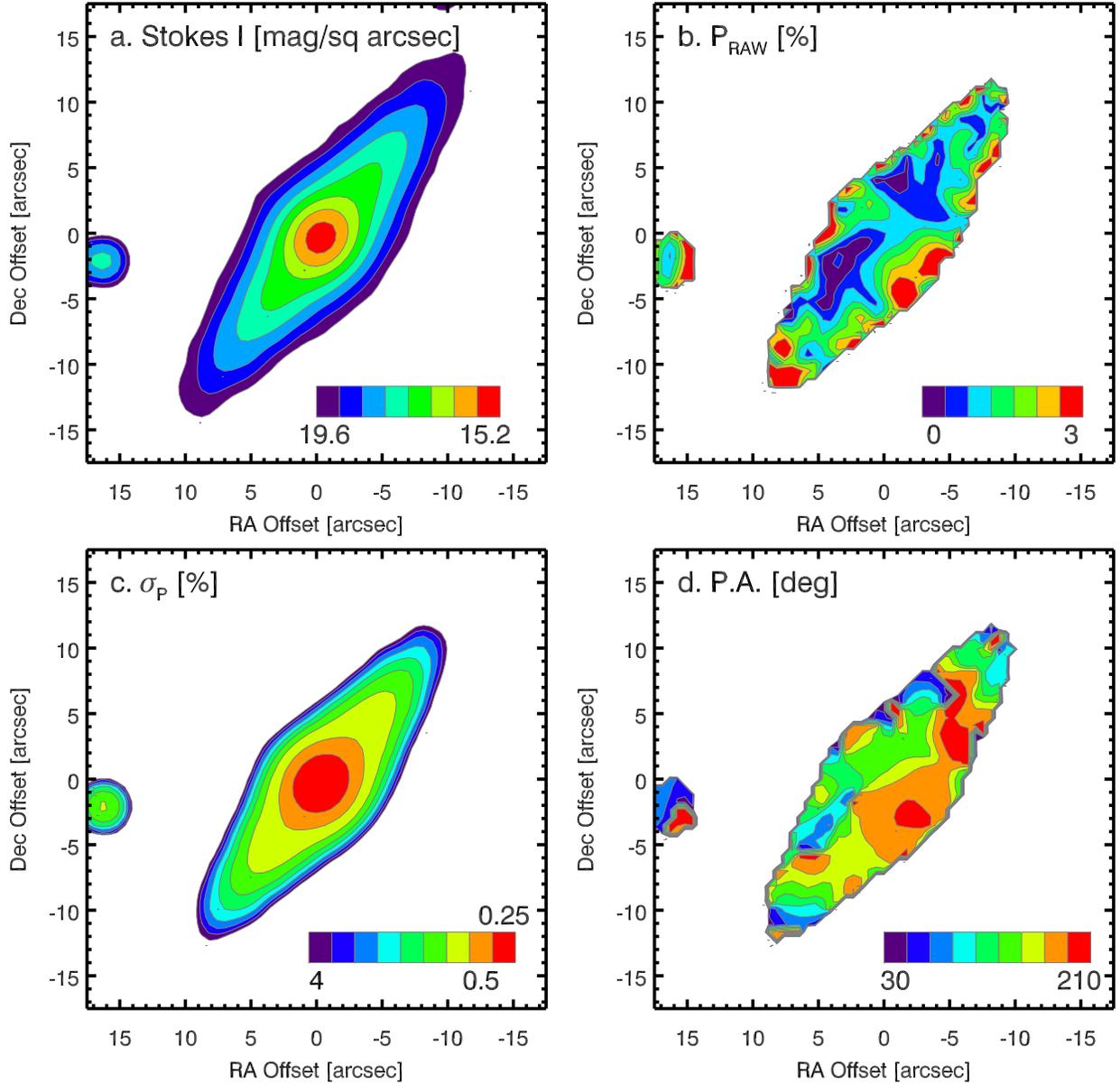}
\caption{Mimir-observed photometric and polarimetric properties of the T-Galaxy.
(a. Top left) Stokes $I$ image of a
$35 \times 35$~arcsec portion of
Figure~\ref{fig_deep_image}, displayed as logarithmically stepped false color contours
of excess surface brightness [in mag arcsec$^{-2}$] above sky. The feature at 
offset (16, $-$2) arcsec is a foreground field star (see Figure \ref{fig_deep_image}).
(b. Top right) Raw percentage polarization $P_{\rm RAW}$, with contour levels stepped
linearly from 0 to 3\%. 
(c. Bottom left) Uncertainty in percentage polarization
$\sigma_{\rm P}$, with contour levels of 0.25\%, then 0.5 through 4\% stepped linearly. 
(d. Bottom right) Observed polarization position angle P.A., measured East from North,
masked outside
$\sigma_{\rm P} \le 4$\%, and displayed as contours stepped linearly from +30\degr\,
to +210\degr.
\newline
\newline
(A color version of this figure is available in the online journal.)
\label{fig_zoom}}
\end{figure}

To polarimetrically analyze the resolved T-Galaxy data, the HWP images 
were instead averaged into four IPPA images 
and combined to produce Stokes $U$ and $Q$ percentage and uncertainty images. 
These were corrected for the Mimir instrumental polarization \citep{Clemens12b},
before uncertainty-weighted smoothing with a  1.75 arcsec FWHM gaussian (matching
the seeing). The resulting $U$ and $Q$ images were used to calculate
raw percentage polarization $P_{\rm RAW}$ and position angle P.A. images, along with 
uncertainties. 
The Stokes $I$, raw polarization percentage $P_{\rm RAW}$, polarization percentage 
uncertainty $\sigma_{\rm{P}}$, 
and polarization P.A. images are all shown in Figure~\ref{fig_zoom}. Each panel in 
the figure spans a  
$35 \times 35$~arcsec portion of the Figure~\ref{fig_deep_image} field
that is centered on the T-Galaxy and includes an unrelated, Galactic star at center left.

The total intensity Stokes $I$ image (Figure~\ref{fig_zoom}.a) shows an inclined, disk-type 
galaxy with a central bulge. Along the major axis, it shows emission 
above the sky level over about 40 arcsec, or about 10 kpc at a distance of 
52.7 Mpc \citep{Crook07}. The color legend in the panel indicates the range of 
excess surface brightness in the $H$ band, above the background of 14.2 mag
arcsec$^{-2}$, spanning 19.6 to 15.2 mag arcsec$^{-2}$ in a logarithmic fashion. 
The polarization percentage uncertainty $\sigma_{\rm P}$ image (Figure~\ref{fig_zoom}.c) shares
the same shape characteristics as the Stokes $I$ image, indicating that poisson
noise dominates the photometric, and 
thereby polarimetric, uncertainties. The $\sigma_{\rm P}$ rendering is clipped 
at an upper value of 4\%, and used as the basis to generate an image mask
that is applied to the $P_{\rm RAW}$ (Fig.~\ref{fig_zoom}.b) and 
P.A. (Fig.~\ref{fig_zoom}.d) images.
These images are displayed as colored contours only between the
limits indicated in the color legends shown at lower right in each panel. 

The $P_{\rm RAW}$ panel (Figure~\ref{fig_zoom}.b) shows polarization
detection across almost the full extent of the galaxy, with lower values
in two zones offset from the center and higher values (albeit more uncertain)
toward the galaxy edges. The P.A. panel (Figure~\ref{fig_zoom}.d) displays
the polarization P.A.s, which have been modified to resolve aliasing about 180\degr, 
so as to track
behavior of the pixels across the histogram of P.A.s (not shown) more smoothly. This results in
apparent P.A. values spanning a range of 30 to 210\degr, as indicated
in the color legend in that panel. The P.A. values seem to favor polarization
orientations of about 130\degr\, to the northeast (NE) of the photometric
center and closer to 188\degr\, to the southwest (SW). The former is 
close to the elongation direction of the major axis (142\degr, see below). Neither is
along the direction favored by the polarization orientations of the stars in the field (Figure~\ref{fig_deep_image}). Thus, although correction for the Taurus foreground
polarization contribution will be important to generating images of $P$ and P.A. intrinsic
to the T-Galaxy, the apparent detections presented in Figure~\ref{fig_zoom}
were not produced {\it by} the foreground polarization.

\subsection{Spectroscopy}

\begin{figure}
\includegraphics[ trim = 2cm 1cm 0cm 1cm, clip=true, scale=0.65]{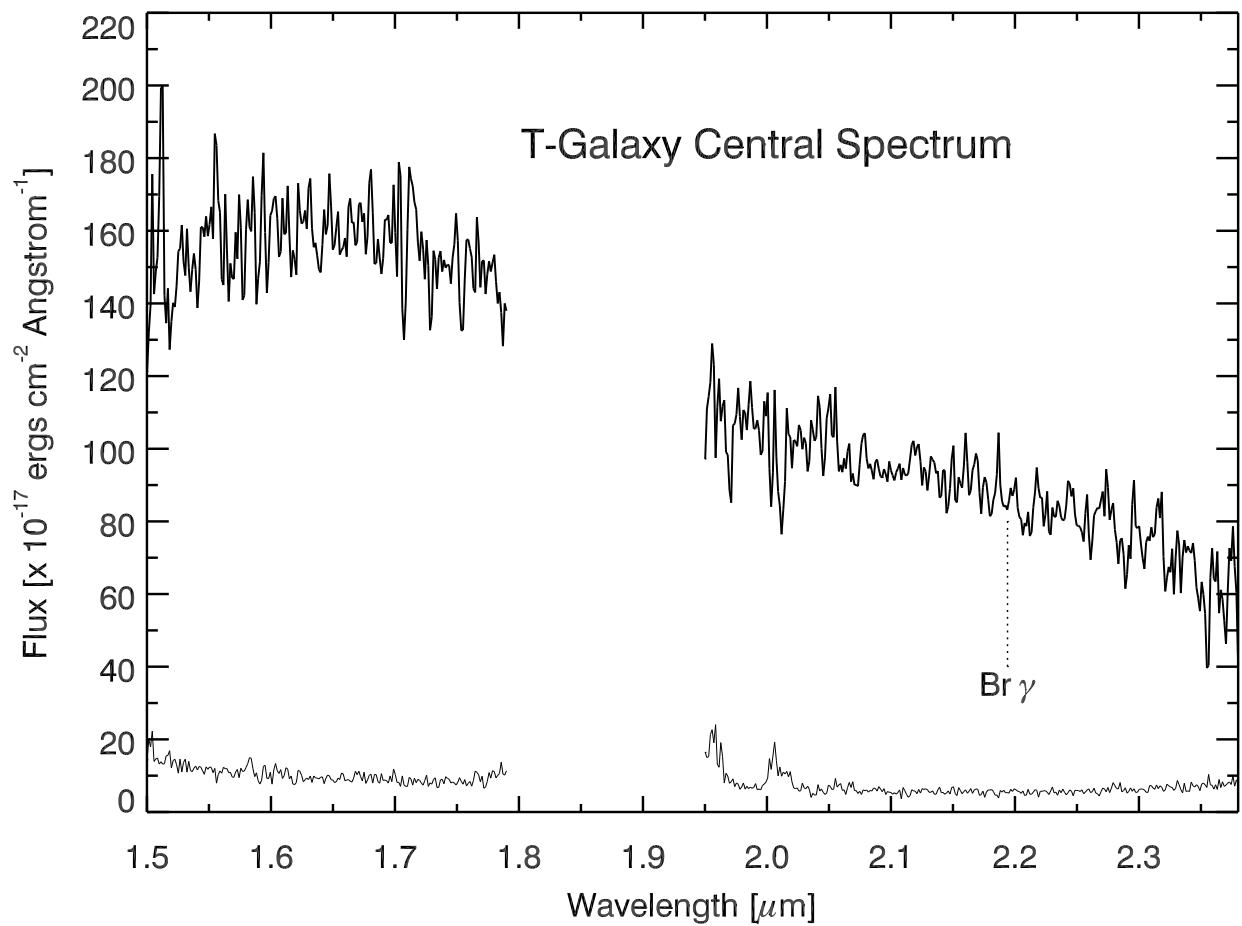}
\caption{Near-infrared spectrum obtained toward the bright core of the T-Galaxy.
Horizontal axis shows wavelength in $\mu$m and vertical axis is in flux units, uncorrected
for slit losses relative to the observations of k Tau. Upper traces show the spectrum
obtained through the $H$ and $K$ band atmospheric windows. Lower traces indicate the
uncertainties. Features at 1.95 and 2.0$\mu$m are due to incomplete atmosphere
removal. The vertical line indicates where the redshifted hydrogen Br-$\gamma$ line
should appear, if present. The lack of strong lines and the overall thermal shape, with
peak near 1.65$\mu$m, argue against the presence of an AGN in the T-Galaxy core.
\label{fig_HK_spec}}
\end{figure}

NIR spectroscopy observations were obtained toward the $m_{\rm H} = 11$ mag
core of the T-Galaxy, using Mimir and the Perkins on the night of 2012 September 25.
The spectral range covered the full H and K bands (1.4 - 2.5$\mu$m) at a
resolution of about 500, using a 1.2 arcsec wide by 5 arcmin tall slit. Eight
spectral images with 5 min exposure time were dithered along the slit. Wavelength
calibration used an Argon lamp. Correction for telluric absorption was done
using observation of k Tau (a bright A0V star) through nearly the same airmass and sky direction.
All spectra were wavelength-calibrated, using the 2D dispersion function obtained
from the Argon images, and registered before adding (with outlier rejection).
The k Tau spectrum was fit and modeled to remove its
hydrogen absorption lines in order to develop a telluric transmission spectrum, using the 
xtellcor routine in the SpeXTool package \citep{Cushing04}. 
This was wavelength matched and divided into the summed T-Galaxy spectrum.
The resulting atmosphere-corrected spectrum for the T-Galaxy is shown as
Figure~\ref{fig_HK_spec}. It presents a mostly 
thermal, cool stellar population appearance, peaking at about 
1.65~$\mu$m. There are no strong emission or
absorption lines seen, to a limit of 5\% of the continuum. This includes a lack of
Br~$\gamma$ (2.166~$\mu$m), either at zero velocity or at the redshift
indicated for the T-Galaxy (see below). 

\section{T-Galaxy General Properties}

This galaxy was first studied by \citet{Crook07}, who used the 2MASS
Extended Source Catalog \citep[XSC:][]{Jarrett00} as input
to the 2MASS Redshift Survey \citep[2MRS:][]{Huchra05a,Huchra05b} 
to search for groups of galaxies containing at least three members. The T-Galaxy was
listed as the second rank member of the five-galaxy group designated number 308
by \citet{Crook07}, whose measured radial velocity of 3868 km~s$^{-1}$ places the T-Galaxy
at a distance of 52.7 Mpc (z $\sim$ 0.013 for $H$ = 73 km s$^{-1}$ Mpc$^{-1}$). 
Its nearest group neighbor is located some 11 arcmin
(170 kpc) away in sky projection, so they are unlikely to be interacting. 

The T-Galaxy Stokes $I$ image (Figure~\ref{fig_zoom}.a) 
exhibits a circular radius encompassing 90\% of the Petrosian flux \citep{Petrosian76}
of 11.75 pixels, or 1.75 kpc at 53 Mpc. The similarly computed 50\% radius lead to
a Petrosian concentration index of 2.9. This 50\% Petrosian radius 
is about twice the FWHM seeing, so the concentration index should be 
unaffected. 

This image of the T-Galaxy 
was fit with a combination \citet{Sersic63, Sersic68}
disk model and an exponential bulge model using the MPFIT package
\citep{Markwardt09}. The galaxy disk was assumed to be circular when viewed face-on
to allow deprojecting the fundamental disk and bulge parameters simultaneously. 
The deprojected bulge was allowed to be elliptical with its own eccentricity and orientation.
The line-of-sight inclination angle was found to be 73.7$\pm$0.1\degr\, (i.e., close to edge-on), 
deprojected along P.A. 141.6$\pm$0.1\degr, measured East from North.

The independent fits to the bulge and disk allowed calculation of the 
bulge-to-total (B/T) flux ratio, found to be 0.24$\pm$0.04. This is a 
brighter bulge than for the Milky Way, whose B/T ratio is in the range 0.11--0.13
\citep{vanderKruit84}.
The T-Galaxy disk scale length was found to be 11.3$\pm$0.3 pixels (1.7 kpc) 
and the bulge scale length was 0.89$\pm$0.03 pixels (130 pc). Milky Way disk
scale lengths are larger, with estimates ranging from 2.26~kpc \citep{Drimmel01} to 
3.3~kpc \citep{Lopez-Corredoira02}.
The T-Galaxy bulge orientation was 142.6$\pm$5.6\degr\, (well-aligned with its disk) 
and the bulge inclination was 60.9$\pm$3.7\degr. 
The radial profile shows a small excess around a deprojected radius of 4 arcsec (1 kpc), 
which could indicate the presence of luminous spiral arms in the disk and 
confirm the spiral galaxy designation. 

The $H$ band magnitude integrated out to the Petrosian radius is 11.35, yielding
an absolute magnitude in this band of $-22.27$ prior to any extinction correction
(see section 4). The radius, concentration index, B/T ratio, and absolute
magnitude are all consistent with the T-Galaxy being a mid-type spiral. The
HK spectroscopy confirms that the 
bulge does not appear to harbor a significant active galactic nucleus (AGN). 

In all of these aspects, the T-Galaxy would seem to be a Milky Way analog, though
smaller, of somewhat lower mass, and with a brighter bulge.

\section{Foreground Extinction and Polarization Determinations}

Because the T-Galaxy is viewed through the nearby \citep[d = 147.6$\pm$0.6 pc:][]{Lionard07}
Taurus Molecular Cloud complex, which is known 
to exhibit both significant extinction, in the form of dark clouds and intercloud material 
\citep[e.g.,][]{Heyer87}, and magnetic fields, as previously traced through optical 
\citep{Moneti84, Heyer87} and NIR starlight polarimetry  \citep{Moneti84, Goodman92,
 Chapman11}, the effects of this foreground material had to be quantified and
removed to obtain the intrinsic T-Galaxy properties. For the foreground
polarization, analysis of the Mimir polarimetry of the stars in the field containing
the T-Galaxy was performed and evaluated. For the photometric extinction and
contributed reddening, analysis of stellar photometry in a somewhat larger field centered on the T-Galaxy
was performed, using several available data sets.

\subsection{Foreground Polarization}

Within the $10 \times 10$ arcmin Mimir field, 239 stars were sufficiently bright
for polarimetric evaluation (i.e., were detected in each of the four IPPAs 
of the HWP).
Of these, twelve showed good polarization signal-to-noise ($P / \sigma_{\rm P} >  2.5$) 
and 227 stars showed lower signal-to-noise. By separating the stars into two stellar samples based
on their S/N ratios, the effects of distance on the foreground polarization contribution could
be examined. This analysis followed that described in \citet{Clemens12c}, 
where stellar sample mean magnitude (which relates to S/N for poisson-limited polarimetry) was
shown to be a useful proxy for distance. The Stokes $U$ and $Q$ percentages 
for each of these
two samples were separately averaged, using inverse variance weighting. 

The same Stokes uncertainty variance weighting was applied to the $H$ band 
magnitudes for each stellar sample,
resulting in a weighted mean of 10.7 mag for the 12 star sample and 13.6 for the remainder
sample. If both samples contain the same types of stars, the
remainder sample would be on average about four times more distant than the 12 star
sample. A high fraction of the stars in both samples are expected to be giants, 
and to be located kiloparsecs beyond \citep{Clemens12a} the much nearer 
Taurus Molecular Cloud complex.
There should be relatively few stars foreground to Taurus in the Mimir field, and any
present would likely show little to no $H$ band polarization.

The Stokes $Q$ and $U$ percentage means for the two samples
were found to agree to within 1.5 times their propagated uncertainties. Thus, 
a significant change in polarization across the distances probed using the stars in the 
field was not found. The two sets
were therefore combined and weighted means were computed for the Stokes percentage parameters.
The weighted mean percentage values found across all 239 stars were:
\begin{eqnarray}
Q_{\rm FG} = 0.81 \pm 0.04 \, \%\, ;\, U_{\rm FG} = 0.75 \pm 0.04 \, \%\,; \\
P_{\rm FG} = 1.11 \pm 0.04 \, \%\, ;\, P.A._{\rm FG} = 21.3 \pm 1.1\degr , \nonumber 
\end{eqnarray}
\noindent
where the subscript `FG' identifies the foreground contribution and  $P_{\rm FG} $ has been
Ricean-corrected to debias for the polarimetric uncertainty, following \citet{Wardle74}.

These parameters best represent the foreground, Galactic
polarization properties (localized in the dust associated with the 
Taurus Cloud complex) along the line of sight to the T-Galaxy. It should be
noted that the portion of the line of sight between the farthest
Galactic star and the T-Galaxy is not represented in our sample of stellar
probes. However, the likelihood of that region having sufficient magnetic field tracing dust 
at these Galactic latitudes to affect the color and polarization of the T-Galaxy is low. 
Alternatively, the Galactic foreground stellar polarizations
could be affected by partially canceling polarization components along the different
pathlengths to the stars. The general coherence of the magnetic
field orientations revealed in Figure~\ref{fig_deep_image} and the
close agreement of the Stokes $U$ and $Q$ values for the two
subsamples make this also unlikely.

Finally, because the foreground polarization is small and the optical depth at $H$ band 
is much less than unity, subtraction of the FG Stokes percentages from the
observed $U$ and $Q$ percentage images for the T-Galaxy is sufficient to correct the polarization properties for the effects of the foreground Taurus layer. 

In summary, the steps taken to develop images of the intrinsic NIR polarization of the 
T-Galaxy begin with forming the four IPPA images from the 16 HWP images and using
these to generate Stokes $U$ and $Q$ percentage images and their uncertainty
images. These were each smoothed, using a gaussian of 1.75 arcsec FWHM with
inverse variance squared weighting from the uncertainty images. From the smoothed
$U$ and $Q$ images, the foreground contributions $U_{\rm FG}$ and $Q_{\rm FG}$
were subtracted. The new $U$ and $Q$ images and their uncertainty images were
combined to form $P_{\rm FG-RAW}$ and P.A.$_{\rm COR}$ images (and uncertainties),
where the subscript `COR' denotes correction for the foreground contributions. Finally, the
polarization percentage uncertainty image was used to Ricean-correct the polarization 
percentage image to yield a $P_{\rm COR}$ image. To allow comparison with radio
synchrotron maps, the Stokes $I$ image was multiplied by the $P_{\rm COR}$
image to produce a polarized intensity ($PI$) image. All of these quantities are 
displayed for the central $22 \times 22$ arcsec of the T-Galaxy in Figure~\ref{fig_p_cor}
and described in detail in section 5.2, below.

\subsection{Foreground Extinction and Reddening}

In addition to correcting for the foreground polarization effects, correcting for the
reddening by the Taurus layer is necessary in order to obtain accurate colors for
the T-Galaxy. In particular, reddening correction in the $(H - K)$ color is needed to
allow comparing the T-Galaxy polarimetry and dust column densities to similar
values for the Milky Way and edge-on galaxies \citep{Jones97}. Such an 
assessment can reveal whether the polarization measured across the T-Galaxy
is dominated by dichroic (selective) extinction, and so able to trace magnetic 
fields, or whether scattering dominates and is unable to reveal magnetic field 
properties.

Several photometric data sets were available for estimating the
extinction and reddening along this sight line, up through the DR7 data release from the
Galactic Cluster Survey (GCS) of the UKIRT Infrared Deep Sky Survey
\citep[UKIDSS:][]{Lawrence07}. Methods for estimating reddening and extinction
from broadband photometric data 
include Near-Infrared Color Excess \citep[NICE:][]{Lada94}, NICE-Revisited
\citep[NICER:][]{Lombardi01}, and Rayleigh-Jeans Color Excess 
\citep[RJCE:][]{Majewski11}. 

The Appendix summarizes the steps, tests, and potential pitfalls associated with
estimating the foreground extinction along this line of sight from these multiple
data sets and methods. The values of $A_{\rm V}$ and $E(H-K)$ finally adopted
were $2.00 \pm 0.10$ mag and $0.125 \pm 0.009$ mag, respectively.

\section{Analysis}

\subsection{2MASS Color and Excess}

The Figure~\ref{fig_deep_image} field was observed here only 
through a single filter ($H$) using Mimir. So, to examine the color variation
across the T-Galaxy,  2MASS images in $J$, $H$, and $K$ bands were 
examined, and an $(H - K)$ color map was formed. This observed color
map was corrected for the foreground color (0.125 mag) and a second
correction of 0.15 mag was subtracted. This latter value is that typically
used in the NICE method \citep{Lada94} as representing the average 
intrinsic color for field stars
(in the absence of a reference field color distribution). The net effect
was the production of a color excess $E(H - K)$ map to use as a proxy for
dust column density across the T-Galaxy.

The distribution of this 2MASS-traced color excess is shown in the upper left
panel of Figure~\ref{fig_E_HK_4}, with its uncertainty shown in the upper right panel.
The $E(H - K)$ panel reveals weak, but significant color shifts across the face of the T-Galaxy. 
Two symmetric arm-like features are seen to the East and West of the galaxy
center as regions of increased reddening, surrounded by regions of decreased
reddening. Using the orientation and inclination properties described in Section 3, a 
deprojected and rotated version of the $E(H - K)$ map was calculated and is presented in the
lower left panel of Figure~\ref{fig_E_HK_4}. Axes for this panel are presented
as linear offsets along the directions of the (projected) major and minor axes, based on the 
distance to the T-galaxy. If the two reddened
arm-like features are spiral arms, they are located only 1 - 2 kpc from the galaxy 
center. They likely correspond to the photometric feature noted at four arcsec radial
offset in Section 3 as possibly indicating the presence of spiral arms. Note that 
this representation only spans the inner 5 kpc where color uncertainties are small;
the full photometric extent is about twice as large.

Finally, the bottom right panel of Figure~\ref{fig_E_HK_4} displays a classification of zones
in the T-Galaxy, selected for their color excess uncertainty, polarization
percentage uncertainty, and, for two of the zones, their color excess values. The selection
criteria are listed in the second through fourth columns of Table~\ref{tbl_p_vs_ehk}.
Classification names of `Core,' `Bulge,' `Interarm,' `Arms,' and `Outer' reflect their
locations and color properties. The first two zones have the smallest uncertainties of 
color excess and polarization, but no color value selection. These criteria map to 
the brightest pixels in the T-Galaxy Stokes $I$ image. The next two zones have 
identical color excess and polarization uncertainty criteria, but were split into 
two groups about $E(H - K) = 0.17$ mag. This produces an `Arms' zone that
encompasses at least the inner, more certain color excess and polarization 
locations showing the redder extinctions identified in Figure~\ref{fig_E_HK_4}.a, as distinct
from the `Interarm' zones that exhibit bluer colors. Finally, the `Outer' zone 
collects the remaining locations for which the polarization uncertainty is under
1.2\%. Analysis of the T-Galaxy classified into these five zones
with the foreground-corrected polarization properties (see next section), enabled
testing for 
correlations between polarization and color excess that may relate to zone classification
and galaxy structure.

\subsection{T-Galaxy Polarization}

\begin{figure}
\includegraphics[ trim = 1cm 1cm 0cm 1cm, clip=true, scale=0.9]{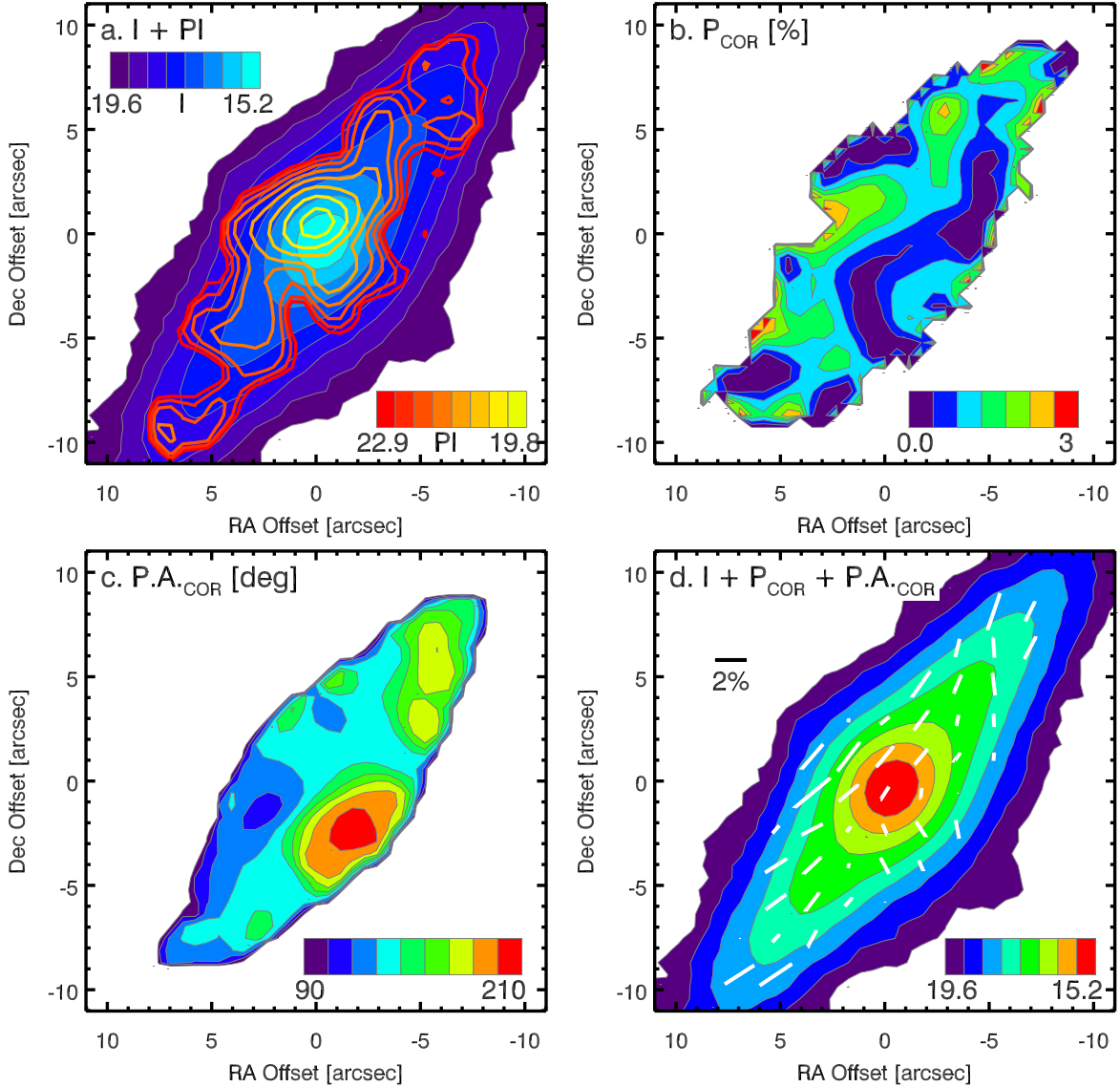}
\caption{Foreground corrected T-Galaxy polarization images over the central 
$22 \times 22$~arcsec region.
(a. Top left) Contours of polarized intensity $PI$ overlaid on a false color 
representation of the Stokes $I$ surface brightness excess [in mag arcsec$^{-2}$] 
above sky, both stepped logarithmically. $PI$ 
contour colors are referenced to the color legend at lower right in the panel; Stokes $I$ 
false colors reference the upper left color legend.
(b. Top right) Foreground and Ricean corrected percentage polarization $P_{\rm COR}$, 
with contours linearly stepped from zero through 3\%, and masked beyond 
$\sigma_{\rm P}$ of 4\%.
(c. Bottom left) Foreground corrected polarization position angle P.A.$_{\rm COR}$, with
contours linearly stepped from  $+90$\degr\, to  +210\degr. 
(d. Bottom right) Stokes $I$ (false color image) with white polarization vector overlays
encoding lengths from $P_{\rm COR}$ and position angles from P.A.$_{\rm COR}$.
\newline
(A color version of this figure is available in the online journal.)
\label{fig_p_cor}}
\end{figure}

\begin{figure}
\includegraphics[ trim = 1cm 1cm 0cm 1cm, clip=true, scale=0.9]{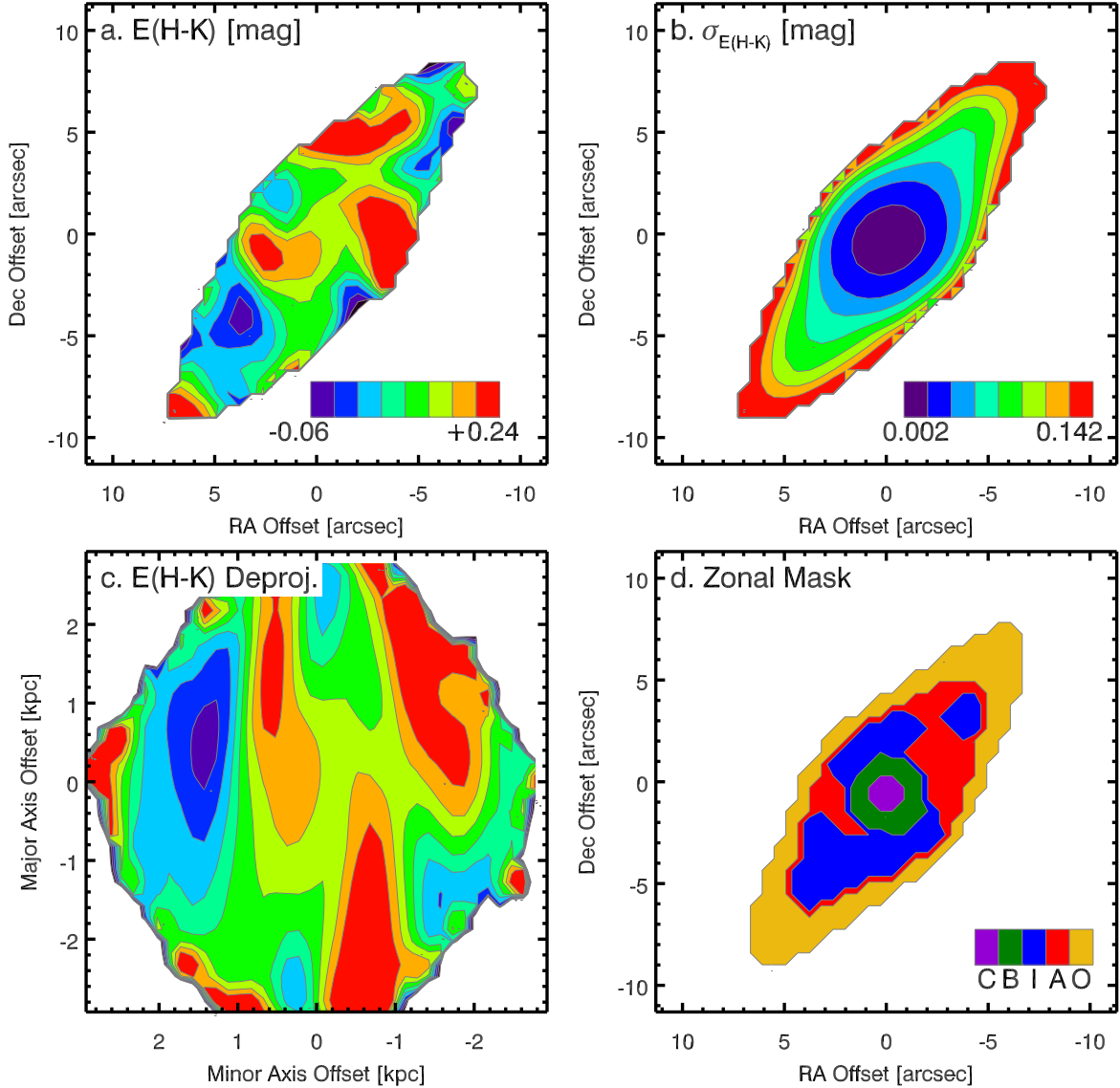}
\caption{The T-Galaxy, as seen in 2MASS $(H - K)$ colors after foreground correction. 
(a. Top Left) Color
Excess $E(H - K)$ map. Note the symmetric arm-like
regions of redder excess colors offset to the East and West of center.
(b. Top Right) Color Excess uncertainty, from 0.002 mag at
center to 0.142 at the edge. 
(c. Bottom Left)
Deprojected and rotated color excess map, showing reddened center, arms and bluer
regions outside the arms. The axes are labeled for linear offset at
the T-Galaxy distance. 
(d. Bottom Right) T-Galaxy zone classifications,
based on colors, color uncertainties, and polarization
uncertainties, as described in Section 5.1 and listed in Table~\ref{tbl_p_vs_ehk}. 
Zone designations, from center outward, are the `Core', `Bulge', `Interarm', `Arms', and 
`Outer'. 
\newline
\newline
(A color version of this figure is available in the online journal.)
\label{fig_E_HK_4}}
\end{figure}

The foreground corrected polarization properties for the T-Galaxy are shown 
for the central $22 \times 22$ arcsec in Figure~\ref{fig_p_cor}. The upper
left panel (Figure~\ref{fig_p_cor}.a) shows contours of polarized intensity
$PI$ overlaid on a false color logarithmic representation of the Stokes $I$
image. There, the brightest peak in $PI$ is offset to the North of the 
photometric center, while a weaker secondary peak is offset to the South.
An alternate explanation is that the photometric peak is bisected by a
depolarization feature that is distributed somewhat linearly and mostly
parallel to the major axis, as though
the bright bulge were shadowed, at least polarimetrically, by the intervening
disk. 
Superposition of disk-parallel and disk-normal (vertical) magnetic fields could also
account for the apparent depolarization.
The $PI$ image shows extensions along the major axis to the NW and
SE, with hints of localized maxima at the ends of the major axis.

The polarization percentage image (Figure~\ref{fig_p_cor}.b) reveals
polarizations of about 1--2\% across most of the extent of the T-Galaxy.
There is a prominent zone of nearly zero polarization percentage surrounding the
isolated 1\% polarized feature centered at an offset of $(3, -3)$ arcsec. This same
feature shows distinctly different P.A. values (Figure~\ref{fig_p_cor}.c)
than the remainder of the T-Galaxy. Most of the T-Galaxy shows
a relatively uniform P.A. of about 130\degr, which is similar to the
major axis position angle on the sky. 

When the corrected polarization percentage and position angle data
are combined with the photometric image, the result, shown in 
Figure~\ref{fig_p_cor}.d, reveals coherent behavior across many
kpc of separation. The vectors shown represent the $P_{\rm COR}$ 
and P.A.$_{\rm COR}$ behavior averaged over $3 \times 3$ pixels 
($\sim$FWHM seeing) and rendered only every three pixels
in each direction to preserve independence. A predominance of disk-parallel polarization
is seen, though in the region SW of the center the
polarization vectors become disk-normal. This SW feature may be 
evidence for a vertical (poloidal) magnetic field or it
could signify scattering of bulge light by disk or halo dust. The
absence of a symmetric zone of disk-normal polarizations to the
NE of the center would seem to favor the latter explanation.

Overall, the NIR polarizations look remarkably similar to many of the radio
synchrotron polarization maps of the disk portions of edge-on
galaxies, such as those in \citet{Sukumar91}, \citet{Dumke95}, 
\citet{Krause06}, and \citet{Soida11}. In particular, the combined
Effelsberg 100 m and VLA radio polarization map of NGC5775
by \citet{Soida11} exhibits virtually all of the same magnetic field
characteristics as seen in the NIR for the T-Galaxy. Both galaxies
show nearly edge-on disks dominated by disk-parallel magnetic fields.
The NGC5775 radio map (Figure 7 of \citet{Soida11}) shows the 
same disk-normal component offset from the center to the SW
as seen here in Figure~\ref{fig_p_cor}.d.

The NGC5775 radio observations are better able to probe fields
in the warm ionized plasma and reveal an `X-shaped' halo field
off of the disk plane (a model is sketched in Figure 12 of
\citet{Soida11}). The NIR of Figure~\ref{fig_p_cor}.d may be
revealing a similar X-shaped halo field to the SW, but some caution
must be applied, as the NIR polarization is more strongly affected
by scattering than would apply to the radio. It is fortuitous that
the VLA observations of NGC5775 and the NIR data reported 
here for the T-Galaxy have about the same effective resolutions 
(about 10-15 independent samples along the luminous disks), 
though NGC5775 is at half the distance of the T-Galaxy and
is twice as large, physically. NIR polarimetry of NGC5775 could
help establish whether the radio X-shaped halo fields are capable
of being revealed in the NIR.

The NIR better traces the
cool, star-forming ISM in the T-Galaxy, revealing a coherent magnetic field 
with a disk-parallel orientation over extents as large as 4 kpc. In addition, there is weak
indication of a rise in polarization percentage and P.A. flaring near the ends of the major
axis that may hint at connections to a more vertical orientation for
a halo magnetic field of the T-Galaxy or a decrease in the disk-parallel
field strength. At present, there are no
radio observations of the T-Galaxy which would permit direct comparison 
with the NIR. Based on the NIR observations, a 
disk-parallel orientation for the B-field is predicted for synchrotron
observations of the T-Galaxy disk at short cm wavelengths.

\subsection{Polarization versus Extinction}

In the disk of the Milky Way, sight lines to reddened stars show infrared
polarization, with the correlation between reddening and polarization
well-modeled using a mixture of uniform and random magnetic field
components distributed into multiple decorrelation cells \citep[][hereafter JKD]{Jones92}.
The polarization arises from the interaction of the magnetic field
with spinning, aligned dust grains, yielding dichroic extinction of the
background starlight.

In the galaxy NGC~4565, \citet[][in his Figure 4]{Jones97} found that points sampled 
mostly across the edge-on dust lane exhibited similar polarization
and color excess behavior to the JKD model, while the similarly edge-on
galaxy NGC~891 showed weaker polarizations. 

The T-Galaxy is less edge-on than
either NGC~4565 or NGC~891. Thus the sight lines through the T-Galaxy
disk will be geometrically shorter, though they could still achieve significant
optical depths. This is true even at $H$ band, because
although the dust extinction per unit gas column is much less than in
the optical, scattering efficiency can be more than twice as high \citep{Wood97b}.

\begin{figure}
\includegraphics[angle=90, trim = 0.7cm 0.7cm 0cm 0.7cm, clip=true, scale=0.8]{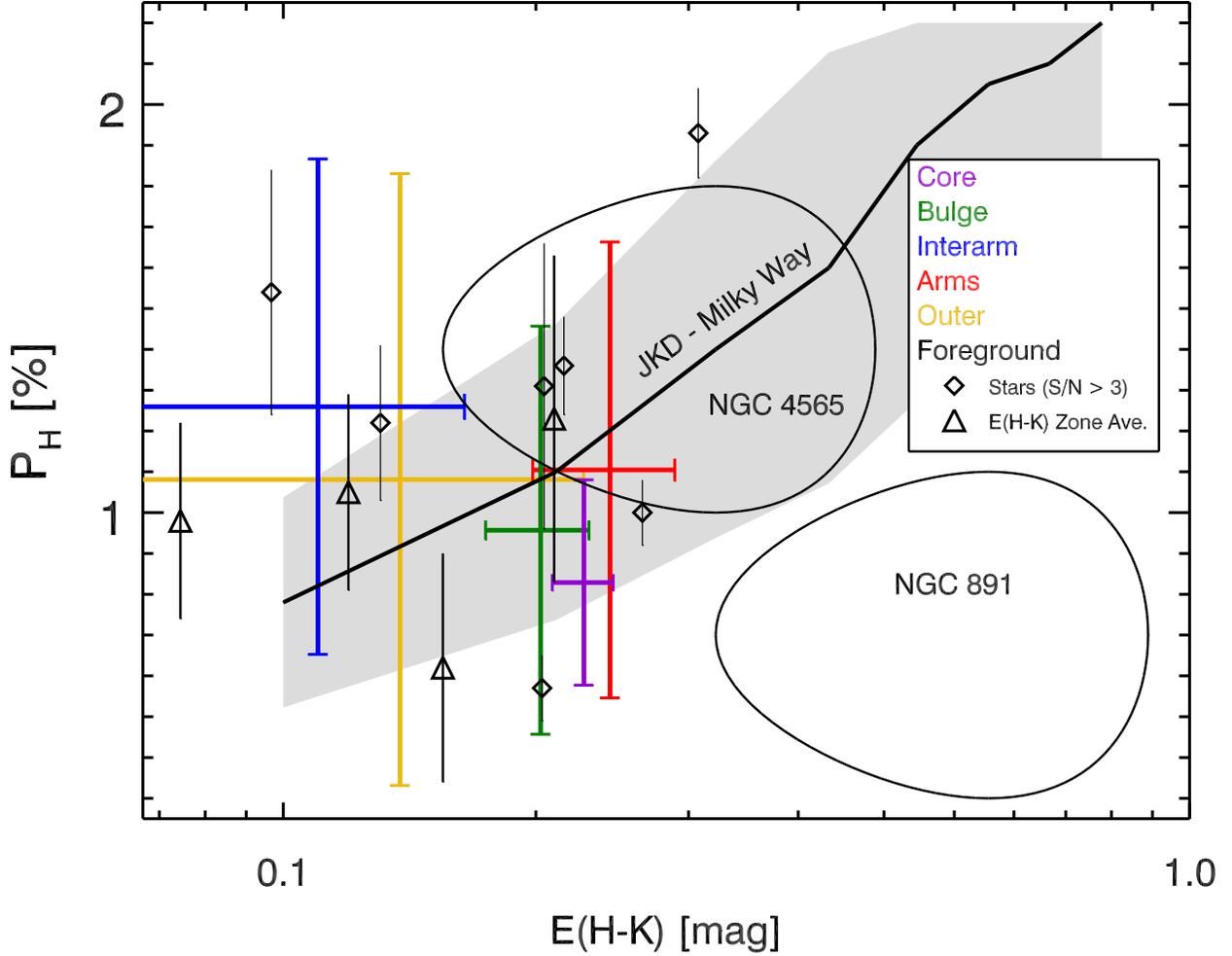}
\caption{Comparison of percentage polarization in $H$ band versus color excess
in $(H - K)$. Colored, crossed error bars locate the T-Galaxy foreground-corrected
zone averages and
$1 \sigma$ dispersions for the five zones defined in Figure~\ref{fig_E_HK_4}.d. 
Black open diamonds are individual Taurus field stars with high polarization significance; black
open triangles are color excess zonal averages of binned low polarization significance stars
(see Section 5.3). Both sets of stars have been reddened and polarized by the
Taurus Molecular Cloud complex.
The solid black curve labeled `JKD - Milky Way' is the approximate behavior
found by \citet{Jones92} toward Galactic stars while the gray band surrounding it
captures 1$\sigma$ deviations of the stars defining the curve. The enclosed regions labeled
`NGC~4565' and `NGC~891' encompass points mapped in these galaxies into 
the $P_{\rm H}$ versus $E(H - K)$ plane by \citet{Jones97}. 
\newline
\newline
(A color version of this figure is available in the online journal.)
\label{fig_p_vs_ehk}}
\end{figure}

The five zone classifications of Figure~\ref{fig_E_HK_4}.d were applied to the T-Galaxy (foreground-corrected) images of polarization percentages and color 
excesses to identify pixel values in each zone. There were
averaged and plotted, along with summaries abstracted
from \citet{Jones97} for NGC~4565 and
NGC~891, in Figure~\ref{fig_p_vs_ehk}. Each of the five T-Galaxy zone characterizations
are shown as colored crossed error bars representing the $1~\sigma$ dispersions
of pixel values of polarization percentage and color excess for each zone. These zone average values and their
uncertainties are also listed in the last two columns of Table~\ref{tbl_p_vs_ehk}.
An abstraction of the JKD curve, based on \citet{Jones97}, is also shown,
surrounded by a gray band depicting the approximate 1~$\sigma$ range of
stellar values seen in Figure 6 of JKD. 

In addition to the T-Galaxy zone averages and dispersions, values of the 
polarizations and color excesses for stars appearing in the Mimir
field of view are shown in Figure~\ref{fig_p_vs_ehk}. Seven of the high signal-to-noise 
polarization stars are
represented as black open diamonds. The remaining, low significance 
polarization stars were grouped into similar NICER-traced extinction bins
and averaged (using Stokes $U$ and $Q$ percentages, with inverse variance weighting). 
These are plotted as black open triangles with vertical error bars.
All of the stellar values reflect the reddening and polarization induced on the
distant stars by the foreground Taurus Molecular Cloud complex.

The T-Galaxy zone averages and sample extents all
fall either within or very close to the region populated by NGC~4565 
measurements and also mostly
fall on or very near the JKD curve. This high degree of agreement of the T-Galaxy
zones with the JKD model strengthens the argument for dichroic extinction and 
a magnetic field origin for the NIR polarizations.

\begin{figure}
\includegraphics[angle=90, trim = 0.6cm 0cm 0cm 0.6cm, clip=true, scale=0.7]{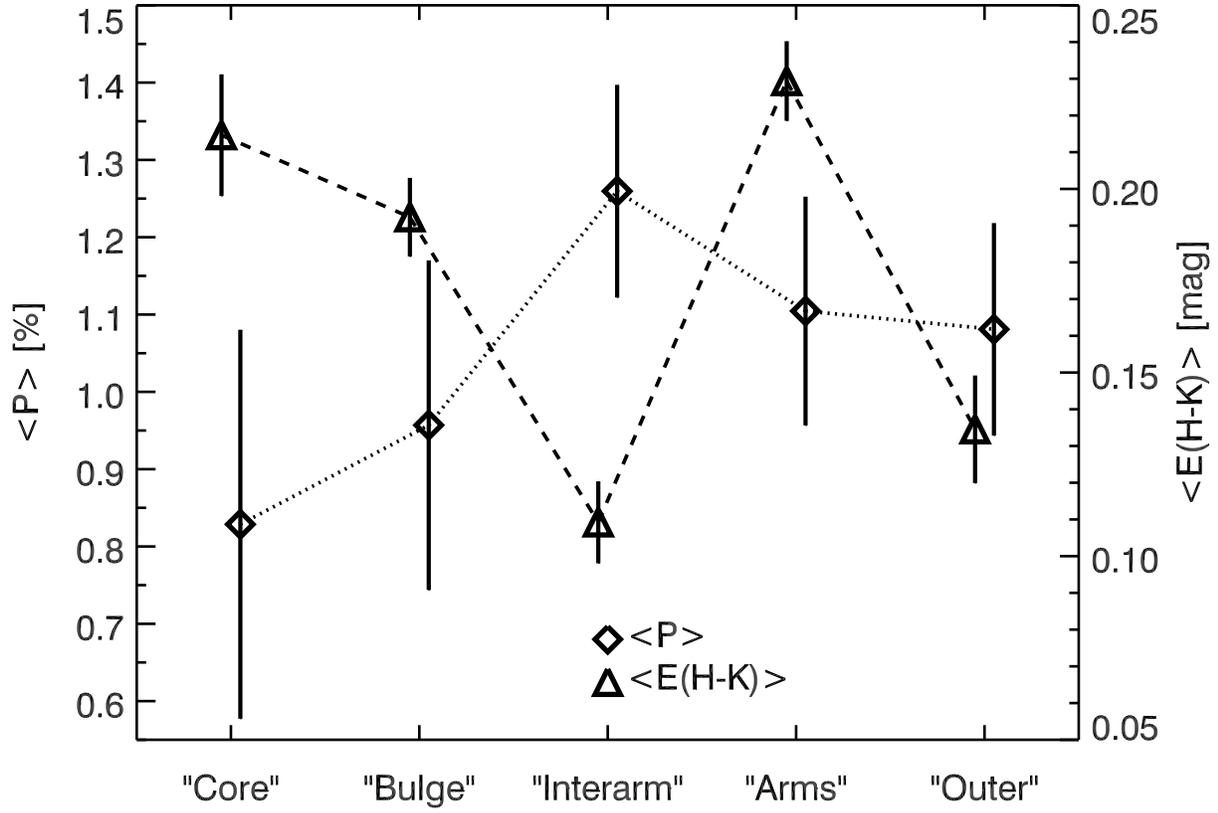}
\caption{Comparison of the zone-averaged percentage polarizations and color
excesses, plotted versus zone in the T-Galaxy. Polarizations are shown as open
diamonds, color excesses as open triangles. Polarizations generally increase from the
Core to the Outer zone. Vertical error bars represent $1 \sigma$ uncertainties of 
the means.
\label{fig_p_vs_ehk_by_region}}
\end{figure}

The mean zone polarizations and color excesses are plotted
against zone designation in Figure~\ref{fig_p_vs_ehk_by_region}. 
The mean polarizations show a weak rise 
from the Core to the Outer zone of the T-Galaxy. The reddest regions 
are the Arms, Core, and Bulge. Though the color excess values
could be due mostly to the zone definitions, the rise in polarization
is not. A similar weak rise in polarization with distance
from a bright central region was noted by \citet{Jones97} for the
case of the nuclear region of IC~694 in Arp~299. He argued that
this was a result of the change from poloidal to toroidal magnetic
field dominance in the polarizing path length. Alternatively, the central
region of the T-Galaxy could be suffering either depolarizing due to
long pathlengths through the intervening disk or a transition from
dichroism in the NE disk to single scattering in the SW portion.

\subsection{Polarization Directions Within the T-Galaxy}

In the modeling of \citet{Wood97b}, such `polarization nulls' near, but 
offset from photometric centers
arise from an admixture of scattering and magnetic field induced
dichroic extinction producing a `crossed-polaroid' effect. This is most
prominent for light originating from the far side of steeply inclined
galaxies. If these effects hold for the T-Galaxy, then the
broken symmetry revealed in the NIR polarimetric
observations may lift the degeneracy on inclination angle and
identify the near and far sides of the T-Galaxy disk. 

The T-Galaxy foreground-corrected polarization P.A. map of Figure~\ref{fig_p_cor}.c 
was examined quantitatively to permit assessing the applicability of the \citet{Wood97a} and
\citet{Wood97b} models. Pixels in the P.A. map were apportioned 
into 25 equal angle azimuthal wedges (relative to the photometric center) 
and were also subdivided into the five zone classifications. 

\begin{figure}
\includegraphics[angle=0, trim = 0cm 0cm 0cm 0.7cm, clip=true, scale=0.55]{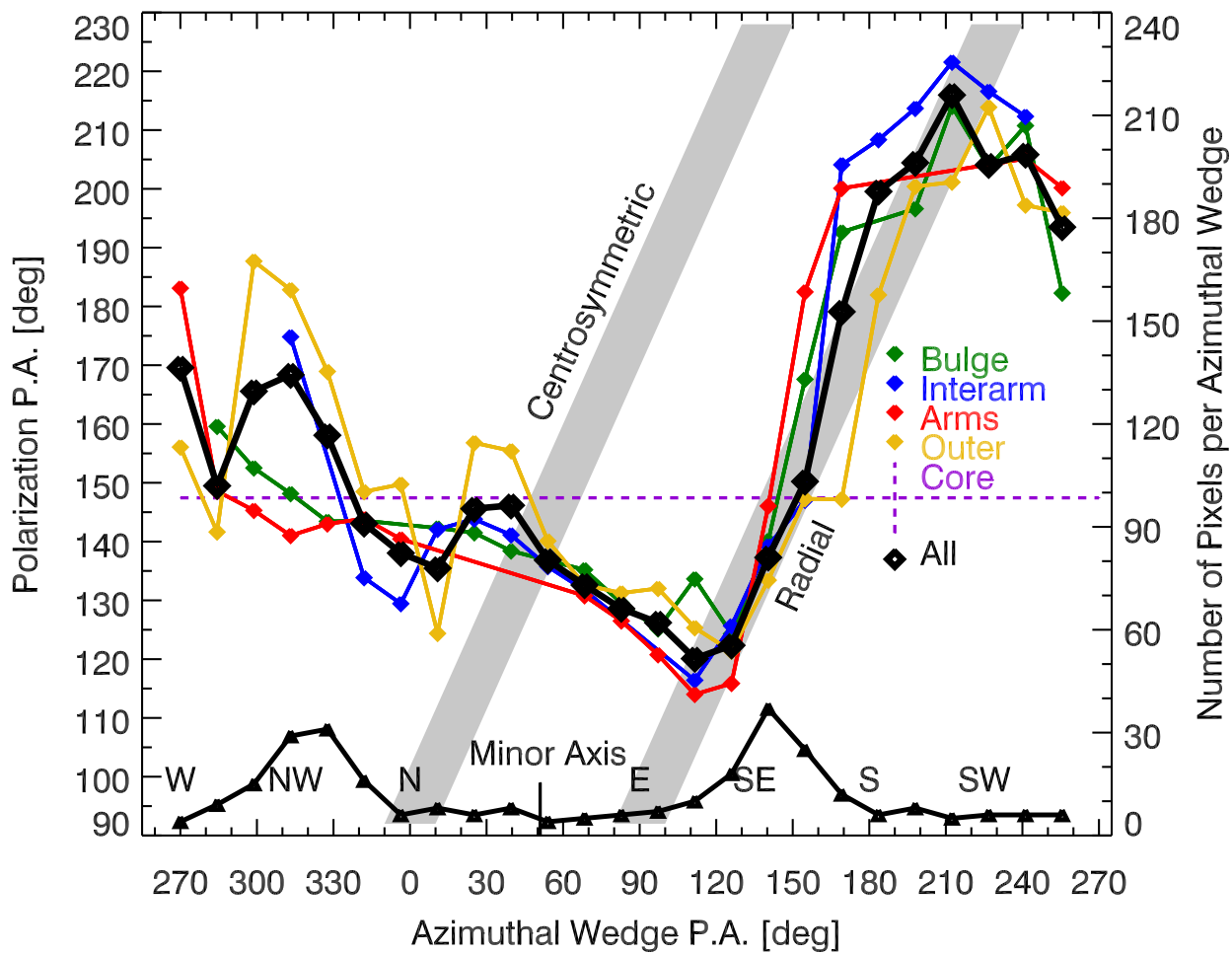}
\caption{Comparison of the polarization position angle, and numbers of pixels, 
versus the position angle for each of 25 azimuthal wedges in the T-Galaxy P.A.$_{\rm COR}$ image, measured 
from the photometric center. Horizontal axis is position angle to the center of each azimuth wedge of
pixels, measured E from N, offset by 270\degr. 
Left vertical axis is pixel averaged polarization P.A.
Right vertical axis is number of pixels in each bin, shown as the lower curve of 
connected black triangles. T-Galaxy major axis is oriented along
142\degr, seen as the increases in pixel numbers near 140\degr\, and 320\degr. 
The polarization P.A. means are plotted for each zone
(as colors) and for all the zones together (black diamonds). 
Core pixels are represented as a dashed violet line and error bar. Two gray diagonal bands 
show behavior for centrosymmetric
scattering and for a radial pattern. 
\newline
\newline
(A color version of this figure is available in the online journal.)
\label{fig_pa_vs_az}}
\end{figure}

Figure~\ref{fig_pa_vs_az} displays how the resulting polarization P.A. and location
are related in the T-Galaxy. Because
the Figure~\ref{fig_p_cor}.c map was not deprojected in this analysis, the number of 
pixels apportioned into each
azimuthal wedge varied. This is shown in Figure~\ref{fig_pa_vs_az} as the
lower, solid black curve and the right hand vertical axis that together
report the number of pixels included in each azimuthal
wedge. The horizontal axis advances North (0\degr), counterclockwise from 
West (270\degr), which is equivalent to the usual `East from North'
meaning with a 270\degr\, horizontal axis shift. The azimuth 
directions to the NW (320\degr) and SE (140\degr) major axis extents of the T-Galaxy 
are seen as increases in the pixel numbers in those bins.

The colored points and connecting lines show the mean polarization P.A. values
for each zone in each azimuth wedge, with the exception of the
Core zone. The Core spans only four pixels, right at the
photometric center, hence it supplies little azimuthal information.
Nevertheless, the Core shows significant polarization, at an average
P.A. of $147$\degr, shown as the dashed violet line and error
bar. Solid lines connect plotted points for the other four zones, 
though azimuth wedges lacking particular zonal pixels were not plotted. The
large black filled diamonds connected by
thick solid black lines represent the all-region average
for each azimuth wedge.

There are a couple of important trends in Figure~\ref{fig_pa_vs_az}.
For the half of the T-Galaxy located roughly North and East of the
major axis, between Azimuth Wedge P.A. values of 270 and 110\degr, there
is a mostly linear, negative slope swing in polarization P.A. from about 160 to
$120$\degr. A very different swing in P.A. is present in the portion of the
T-Galaxy located South and West of the major axis, where
the polarization P.A. values jump up to a mostly constant value near +200\degr.

Two gray zones in the Figure highlight the behaviors expected for a 
centrosymmetric (scattering) pattern and for a radial pattern, which
can also arise from scattering, though not from a single central source
\citep{Wood97b}. None of the P.A. values line up with the centrosymmetric
zone, ruling out strong centrally-illuminated single scattering dominating the NIR polarization
pattern for the T-Galaxy. 

The SW half of the T-Galaxy shows some of 
the weakest polarization percentages (Figure~\ref{fig_p_cor}.b)
with a region-averaged $P_{\rm COR}$ of $0.4 \pm 0.1$\%, while the NE 
half averages $1.4 \pm 0.1$\%.
The zone of low SW polarization and most radial polarization P.A. orientations
may be evidence for the `crossed polaroid' effect
described in \citet{Wood97b} for the far-side of inclined disks, though
mostly affecting only a portion of the SW half.
The near-side of the disk, here the NE half, better
permits starlight to experience magnetic field
tracing dichroic extinction, leading to the observed polarization,
while the far-side (SW) suffers depolarization and/or single scattering.

There is remarkably little difference in the polarization P.A. 
azimuth wedge averages for the different zones in the T-Galaxy.
The Outer zone may exhibit somewhat more angular scatter, but
this is likely a result of the higher polarization P.A. uncertainties there.
The strong agreement of the polarization P.A.s across the 
many zones and across almost the full extent 
of the T-Galaxy is strong evidence for a large-scale uniform magnetic field
component.

\section{Discussion}

In exploiting the opportunity provided by the serendipitous NIR polarimetric
observation of the T-Galaxy, as augmented by analyses of other existing 
stellar data, have clear understandings emerged regarding the
nature of the T-Galaxy and how its properties might relate to those in the Milky Way
and other normal galaxies?

First, is the T-Galaxy a Milky Way analog? Is it a disk galaxy that lacks
a central AGN (so that the polarization observed is not AGN related)? 
The decomposition of the Stokes $I$ image into bulge and
disk components argues for a significant bulge contribution to the total
luminosity, but not one so high as to be AGN dominated. The B/T ratio
and Petrosian concentration index are both comparable to values seen
for other disk galaxies \citep[e.g.,][]{Benson07, Lackner12}, as are the 
surface brightness, disk scale length, and bulge scale length \citep{MacArthur03, Ganda09}.
Optical spectroscopy of the nucleus could test for the presence of an AGN,
though \citet{Crook07} do not comment on this. Our own HK-band
spectroscopy shows a cool stellar population (thermal-like) spectrum, that is 
unchanged when deextincted. Further, the lack of narrow or broad emission lines 
in our HK spectrum argues
against the presence of a strong AGN. A fairly conservative upper
limit of 0.15\% of the T-Galaxy luminosity for its nuclear emission lines
was arrived at through combining three times the noise in the spectrum 
with a notional linewidth of 150 km s$^{-1}$. Such a low upper limit on the
spectral line power for the nucleus of the T-Galaxy rules out normal Seyferts and
other AGNs.

The bulge to disk luminosity ratio is consistent with 
Sb-Sbc types \citep{Graham08}. A bump at four arcsec radial offset in the photometric 
profile of the Mimir $H$ band image used in the S{\'e}rsic fits hinted at
possible spiral arms. The 2MASS $(H - K)$ color excess image, after
foreground reddening correction, showed evidence for symmetric
red features emanating from the vicinity of the core/bulge and 
extending to the T-Galaxy edges. These red features are likely to be
tracing the dust enhancements expected for spiral arms in the T-Galaxy
and they appear at the right radial offset to match the photometric
bump. The 2.3 kpc circular radius to 90\% of the Petrosian flux is 
significantly smaller than the extent of the Milky Way disk, but 
other aspects indicate that the T-Galaxy is a fair, though smaller, 
Milky Way analog.

Second, have magnetic fields in the T-Galaxy been detected? Polarization
can arise from many sources, though in the NIR, generally only scattering
and dichroic extinction will dominate in normal galaxies. Modeling by
\citet{Wood97a} and \citet{Wood97b} of NIR polarization from steeply
inclined disk galaxies explored the observational effects of each of 
scattering and dichroic extinction alone, as well as together. When 
scattering dominates, \citet{Wood97a} found that face-on galaxies show
strong centrosymmetric polarization patterns and edge-on galaxies
show polarizations perpendicular to the disk. Dichroic extinction tends
to produce polarizations parallel to the disk, since the
disk magnetic field they modeled was toroidal. The model of mixed scattering 
plus dichroism
for an inclined galaxy, shown as Figure~3 (center) in \citet{Wood97a},
contains many of the polarization features and properties present
in Figure~\ref{fig_p_cor}.d here. 

The azimuthal variation of polarization P.A.s, shown here as Figure~\ref{fig_pa_vs_az},
reveals both the negative-sloped swing of P.A. for the higher mean polarization
(NE; near) side of the T-Galaxy disk
as well as the strong positive unity-sloped P.A. swing that signifies radial (single
scattering) for the lower mean polarization
(SW; far) side of the disk. 
This azimuthal variation is also seen in the radio polarization map of 
NGC5775 \citep{Soida11}.
Similar 
quantitative characterizations in future modeling studies could help
sharpen comparisons and interpretations. This would also address
whether NIR and radio polarizations are both probing halo magnetic
fields equally or if the NIR is affected by scattering.

The overall match of the T-Galaxy polarization percentages and
color excesses with the Milky Way based JKD model and with the
values seen toward the edge-on galaxy NGC 4565, shown in 
Figure~\ref{fig_p_vs_ehk}, argue for a common magnetic field plus
dichroic extinction origin for the polarizations and reddenings from these
three galaxies. 

At present, the T-Galaxy lacks resolved kinematic observations, which are needed
to ascertain the rotation direction. The polarization map does reveal a broken symmetry
about the major axis, which interpreted under current models was used to
predict the near and far-side locations of the T-Galaxy disk. The same map
also showed coherent P.A. swings. Kinematic
data would be able to test these near-side/far-side predictions.

This galaxy was not targeted, but was contained in a field of deep Taurus
stellar observations, and these stellar values were able to reveal and
calibrate the foreground polarization properties. The stellar 
polarization properties show the orientation of the local Galactic magnetic field 
contained within the Taurus layer and
could be combined with observations from {\it Planck}
to improve the removal of Galactic foregrounds from CMBR observations
aimed at mapping B modes \citep{Jones03}.

\section{Summary}

Deep $H$ band linear polarization observations toward one
$10 \times 10$ arcmin field in Taurus serendipitously contained an
inclined, 53 Mpc distant, (small) Milky Way analog galaxy. The
polarization data were sensitive enough to permit measuring the
intrinsic polarization generated by the combination of magnetic
fields and aligned dust grains across nearly the full extent of that galaxy.

Because the galaxy was observed {\it through} the gas and
dust in the Taurus Molecular Cloud complex, it was necessary to
measure and remove the foreground contributions by Taurus to the distant
galaxy polarization, extinction, and reddening. The 
foreground extinction and reddening were measured using multiple 
methods applied to several stellar photometric data sets. Corrected for this reddening, 
an $E(H-K)$ color excess map of the galaxy revealed red spiral arms 
at locations in agreement with a photometric bump in the radial $H$ band
profile. 

Once foreground-corrected, the distant galaxy NIR polarizations were 
found to be comparable to Milky Way stellar values, indicating that
they most likely are tracing the uniform component of the magnetic
field in the galaxy disk. The field revealed shows remarkable similarities
to the magnetic fields revealed by radio synchrotron polarimetry of edge-on
disk galaxies. As NIR polarimetry for this galaxy mostly
arises from dichroic extinction in the cool, star-forming interstellar
medium and radio synchrotron arises in a much warmer component,
the correspondence of the magnetic field patterns for the two favors a single uniform magnetic
field threading both components.  

A broken symmetry in the degree of 
polarization and its orientation direction with respect to the 
major axis was found, likely due to a change
in the relative mix of dichroic absorption and scattering from 
the NE front and SW back sides of the disk of that galaxy.

The approach of using the stellar foreground polarization properties
to characterize the local, Galactic magnetic field projected orientation
could prove useful for constraining models aiming to correct {\it Planck}
CMBR observations for Galactic contamination.

\acknowledgments

This publication makes use of data products from: the Two Micron All Sky Survey, 
which is a joint project of the University of Massachusetts and the Infrared 
Processing and Analysis Center (IPAC)/California Institute of Technology (CalTech), funded by 
NASA and NSF;  the NASA/IPAC Extragalactic 
Database (NED), which is operated by the Jet Propulsion Laboratory (JPL), 
CalTech, under contract with NASA;  the {\it Wide-field Infrared Survey Explorer}, 
which is a joint project of the University of California, Los Angeles, and 
JPL/CalTech, funded by NASA;  observations made with the {\it Spitzer Space Telescope}, which is operated by JPL/Caltech, under a contract with NASA;
the UKIDSS project, as defined in \citet{Lawrence07}, and which used the UKIRT Wide Field Camera \citep[WFCAM:][]{Casali07}, the photometric system described in \citet{Hewett06}, 
the calibration described in \citet{Hodgkin09}, and the pipeline processing and science archive  
described in \citet{Hambly08}. 
This research made use of the C. Beaumont IDL library, available at
www.ifa.hawaii.edu/users/beaumont/code. Careful reading of the manuscript
by A. A. West and useful comments by the anonymous reviewer are appreciated.
This research was conducted in part 
using the Mimir instrument, jointly developed at Boston University and Lowell 
Observatory and supported by NASA, NSF, and the W.M. Keck Foundation.
Grant AST 09-07790 from NSF to Boston University and grants of 
observing time from the Boston University 
-- Lowell Observatory partnership are gratefully acknowledged.

{\it Facility:} \facility{Perkins}

\appendix

\section{Foreground Extinction and Color Excess Determination}

The NASA/IPAC Extragalactic Database (NED) reported results from the 
model of \citet{Schlegel98}, based on {\it COBE}/DIRBE and {\it IRAS}/ISSA
image analysis, as updated using Sloan Digital Sky Survey (SDSS) data 
by \citet{Schlafly11}, for estimating foreground extinctions toward the T-Galaxy.
These had an effective angular resolution of about 6 arcmin \citep{Schlegel98},
comparable to the entire Mimir field of view. The values reported for
the \citet{Schlafly11} update are $A_{\rm V} = 4.127$ mag and 
$A_{\rm H} = 0.676$ mag, the former is listed in the first row of Table~\ref{tbl_ext}
(the original \citealt{Schlegel98} values are listed in italics in the second row
of the Table). When converted to reddening, using \citet{Indebetouw05}, 
the Taurus layer contribution would have been 
$E(H - K) = 0.26$ mag. The overall observed T-Galaxy $(H - K)$ color
from 2MASS is 0.42 mag, as reported in NED (and was confirmed here by
differential photometry in the images, tied to the colors of eleven 2MASS PSC 
stars located within two arcmin of the T-Galaxy). Hence, correction of the observed 
color by this indicated
reddening would yield a galaxy $(H - K)$ color of 0.16 mag, which is significantly bluer than
the colors of normal spiral galaxies \citep{Mannucci01}. Indeed, a quick search
of SDSS galaxies reveals only 3.3\% show bluer colors (A. A. West, priv. comm.). This blue galaxy
conclusion seemed unlikely, so several other methods and data sets were examined
and the initial two rows of Table~\ref{tbl_ext} were not used in the subsequent
analyses.

The direction to the T-Galaxy misses the most highly extincted Taurus clouds
(the direction is near and between the Eastern ends of TMC-1/HCL-2 and TMC-2/B18)
showing the strongest CO emission \citep{Goldsmith08} in the region, 
and thus is expected to be a low-density, low-column ISM line of sight. 
The \citet{Goldsmith08} maps suggest a total H$_2$ column density 
between about $2 - 2.5 \times 10^{21}$ cm$^{-2}$ (A$_{\rm V} = 2 - 2.5$ mag;
row 3 of Table~\ref{tbl_ext}). This extinction range is also indicated in the
NICER-based map, derived from 2MASS $JHK$ photometry, presented by \citet{Lombardi10}
and listed as the fourth row of Table~\ref{tbl_ext}.

To obtain better extinction and reddening estimates, catalog data from
several surveys were mined and analyzed for a region centered on the 
T-Galaxy and also for a 
reference region located 1.9\degr\, away ($\alpha$[J2000] = $71.{\degr}125$;
$\delta$[J2000] = $+22.{\degr}833$).  The reference region was 
chosen, based on examination of the \citet{Goldsmith08} and \citet{Lombardi10} 
maps, as a direction relatively near on the sky to the T-Galaxy and 
showing no associated Taurus cloud CO emission or 2MASS extinction.
Stellar photometry catalogs for both the T-Galaxy zone and the reference zone were searched
out to 10 arcmin radii from those centers. The catalogs included 2MASS,
UKIDSS, the {\it Spitzer Space Telescope} \citep{Werner04} archive, and the {\it WISE} 
\citep{Wright10} All-Sky Data Release products. 

The resulting point source catalogs were position-matched and merged with the Mimir
polarimetry and $H$ band photometry several ways to test the effects of 
inclusion of the various catalogs. For example, the {\it Spitzer} archive included deep
InfraRed Array Camera \citep[IRAC:][]{Fazio04} observations for a Taurus `Legacy-Class' 
survey \citep{Padgett08}
that overlapped the T-Galaxy field, but did not overlap the
reference field. The {\it WISE} all sky survey
observed both fields, but had higher uncertainties
in its W2 (4.6~$\mu$m) band, compared to B2 (4.5~$\mu$m) for IRAC. 
The best photometry toward the
T-Galaxy field would include the {\it Spitzer} data, but such a choice could 
introduce bias. The {\it Spitzer} IRAC/B2 magnitudes for stars in the T-Galaxy field
were compared to the matching {\it WISE}/W2 magnitudes and found to be
in excellent agreement (mean difference = $0.011 \pm 0.012$ mag for 374
stars), indicating that use of {\it Spitzer} for the T-Galaxy field and {\it WISE}
for the reference field would not introduce meaningful bias.

Use of UKIDSS (only the $K_{\rm 1}$ band data for this region were
contained in DR7) and Mimir deep $H$ photometry added fainter stars beyond the 2MASS
sensitivity limits, though most of these did not have corresponding {\it Spitzer}
or {\it WISE} photometry. Given the weak reddening seen for {\it Spitzer} 
IRAC bands \citep{Indebetouw05}, compared to the $J$, $H$, and $K$ bands, all
{\it WISE}/W1 (3.4~$\mu$m) and {\it Spitzer}/B1 (3.6~$\mu$m) magnitudes 
were considered to be equivalent
$L^\prime$ (3.45~$\mu$m) band values, and similarly the W2 and B2 
magnitudes were considered to be 
equivalent $M^\prime$ (4.75~$\mu$m) band values, despite the somewhat
different bandpasses and mean
wavelengths for the bands used for the two missions. The combined data sets for 
each region were assembled using the lowest uncertainty photometric value 
for each star for each band (with sufficient departure tests from this requirement
to ascertain a lack of important biases).

The three main methods for obtaining extinctions and reddenings from
these merged data sets used the data in different ways. The NICE method
relies only on $H$ and $K$ bands, and infers extinction from the reddening
in the $(H - K)$ color relative either to means for unextincted stars, or to
means of stars in an unextincted reference field. Here, the 581 stars with
$(H - K)$ colors in the T-Galaxy region yielded an average A$_{\rm V}$ of
$1.66 \pm 0.12$ mag. A value of $-0.14 \pm 0.12$
mag was found for the 475 stars in the reference field. Both values were
obtained when each
region's stars were compared to an average intrinsic color $(H - K)_0$ of 0.15 mag
\citep{Lada94}. The near zero extinction for
the reference field obtained for the NICE method would appear to be 
confirmation of that method's utility. Differencing the T-Galaxy and reference
zone values yielded the
A$_{\rm V}$ listed in the fourth row of Table~\ref{tbl_ext}.

The NICER method\footnote[3]{Using the C. Beaumont IDL implementation}
utilizes either all three 2MASS bands ($J$, $H$, and $K$),
or can be extended to include any number of bands. Applying NICER/3 ($J$, $H$, 
and $K$ bands only)
to the reference region returned $A_{\rm V} = 0.71 \pm 0.08$~mag, in sharp
contrast to the near zero value from NICE. 
 Including the $L^\prime$ and $M^\prime$ band magnitudes (i.e., NICER/5) increased
this to $0.88 \pm 0.09$ mag, increasing the concern that NICE may be in error.

The RJCE method is based on $(H - M^\prime)$ colors, as the intrinsic spread of
such colors for unextincted stars is exceedingly small \citep{Majewski11}, much smaller than
for the $(H - K)$ color used by NICE. Application of RJCE to the reference
region yielded A$_{\rm V} = 0.61 \pm 0.19$ mag, not far from the NICER values.

Hence, the reference region, though selected to avoid the
Taurus dark and molecular clouds, and although showing no extinction
under the NICE method, exhibits significant NICER and RJCE traced near-infrared
color and therefore dust extinction,
most probably associated with atomic or ionized hydrogen gas and not
molecular gas. Examination of $WISE$ W3 (12~$\mu$m) images revealed the
presence of extended dust emission, in the form of IR cirrus, confirming the
stellar color findings from NICER and RJCE.

A second reference region, previously studied by \citet{Swift08}, was selected
to serve as the zero extinction calibrator. While still within the large Taurus
region, it is further on the sky from the T-Galaxy than the first reference
region. This may affect comparison of the stellar types present in this second
reference region and the T-Galaxy field stars. Similar to the approach already
described, existing photometric databases were mined for photometry in the
five NIR bands. This new region, similar to the first reference region, had
$WISE$ data but no $Spitzer$ data.
The stellar colors, uncertainties, and covariances needed to calibrate the 
NICER method \citep{Lombardi01} were obtained for the 261 stars located within
10 arcmin of the second reference center ($\alpha$[J2000] = $65{\degr}.00$,
$\delta$[J2000] = $17{\degr}.00$) and having low uncertainties ($\le 0.07$ mag)
for all colors. 
Using this second reference zone calibration, the T-Galaxy region stellar 
data set was analyzed using NICER/3 (i.e., $JHK$),
NICER/5 ($JHKL^\prime$M$^\prime$), and RJCE, with all results listed in 
the final three rows of 
Table~\ref{tbl_ext}. 

\begin{figure}
\includegraphics[angle=0, trim = 2.5cm 0.5cm 0cm 0cm, clip=true, scale=0.65]{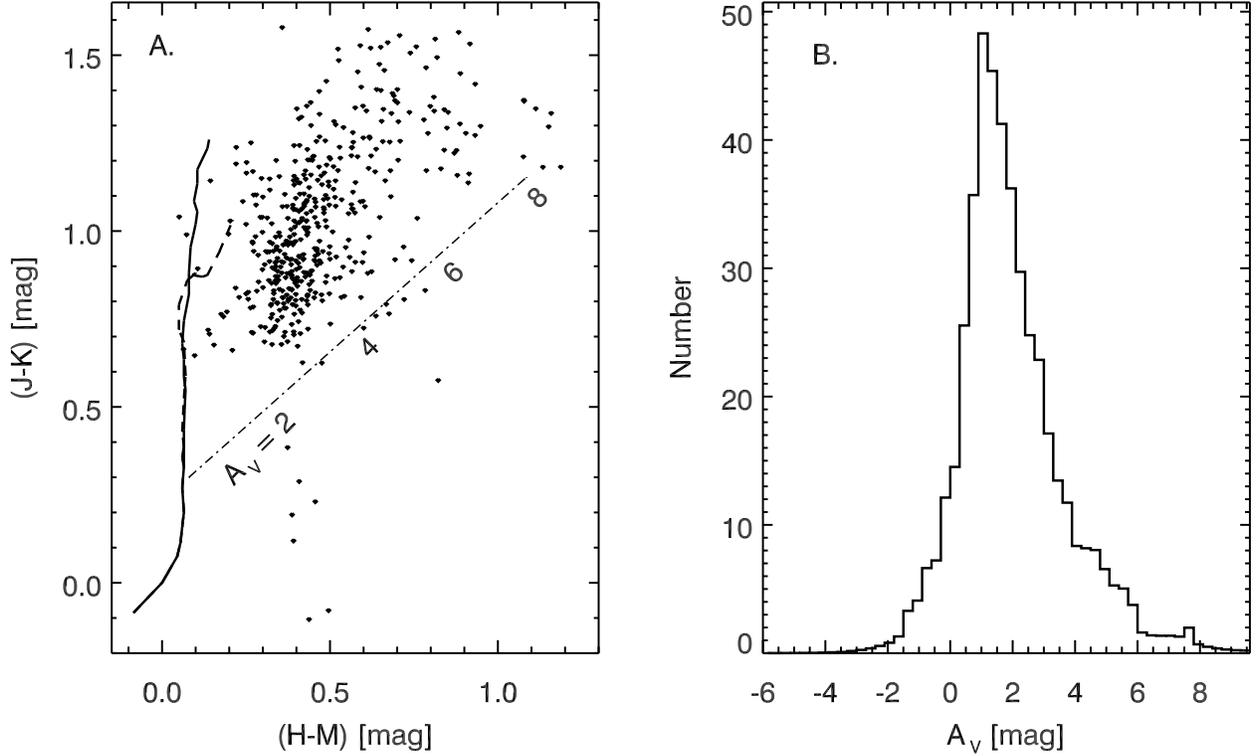}
\caption{Extinction and reddening of stars in the T-Galaxy region, as revealed using the RJCE method.
(a. Left Panel) Color-color diagram for 358 stars located within 10 arcmin of the
T-Galaxy and having $JHKL^{\prime}M^{\prime}$ magnitudes. Horizontal $(H - M^\prime)$ color is used in the RJCE
method \citep{Majewski11} for determining extinctions. Vertical axis shows $(J - K)$ color.
The solid and dashed curves are the \citet{Bessell88} loci of colors for unextincted 
giants and dwarfs, respectively. The \citet{Indebetouw05} reddening 
vector is shown as the angled, dot-dashed line,
with $A_{\rm V}$ values (in mag) shown. The stars in the field surrounding the T-Galaxy
exhibit significant extinction. 
(b. Right Panel) Histogram of RJCE extinctions determined
from the stars shown in the left panel. Note the lack of any second strong extinction population that
would indicate a second extincting layer. The direction to the T-Galaxy appears to pass 
through extinction associated only with the Taurus Molecular Cloud complex at
about 150 pc distance.
\label{fig_rjce}}
\end{figure}

Figure~\ref{fig_rjce} presents an illustration of the application of the RJCE method.
In the left panel, the color-color
distribution of $(J - K)$ versus $(H - M^\prime)$ stellar values for the T-Galaxy field 
are presented and,
in the right panel, the distribution of A$_{\rm V}$ values derived from these
data using RJCE are shown. The solid and dashed black lines in the left panel show the \citet{Bessell88}
loci for unextincted giants and dwarfs, respectively. The near vertical run between
0.1 to 1.2 mag of $(J - K)$  highlights the narrow spread in intrinsic 
$(H - M^\prime)$ colors exploited by the RJCE method. 

To compute the A$_{\rm V}$ histogram shown in Figure~\ref{fig_rjce}.b, 
the measured photometric
uncertainties were propagated to extinction uncertainties.
These uncertainties were typically around 1 - 1.5 mag, much larger than 
the bin widths typically used when constructing A$_{\rm V}$ histograms.
To build this histogram, each A$_{\rm V}$ value and
its uncertainty were replaced with a set of several thousand, normally
distributed, Monte Carlo 
realizations and these were placed into the corresponding histogram bins.
Without this probabilistic approach, A$_{\rm V}$ histograms become artificially 
noisy and subject to over-interpretation. The A$_{\rm V}$ histogram 
in Figure~\ref{fig_rjce}.b shows that the
direction to the T-Galaxy is well-represented by a single extinction
and polarization layer, localized to the Taurus Molecular Cloud complex,
with little to no extinction beyond Taurus to the multi-kpc distances of the Galactic
stars probed.

Taking into account the several methods used for analyzing the stellar
data sets, values of A$_{\rm V} = 2.00 \pm 0.10$ mag and
 $E(H - K) = 0.125 \pm 0.009$ mag
(Table~\ref{tbl_ext}) appear to be the best characterizations of the line of sight to
the T-Galaxy. The most conservative correction might consider only the stellar
difference values between the T-Galaxy region and first reference region examined, but,
as described above, that reference region (and by inference most of 
the Taurus regions mapped in CO and using NICER) seems to exhibit
an extinction nearly as large as that found for the T-Galaxy direction. This 
required use of the second, more angularly offset reference region of
\citet{Swift08}. The even higher extinctions found by 
\citet{Schlegel98} and \citet{Schlafly11} through
combining $COBE$, $IRAS$, and SDSS data (first two rows of Table~\ref{tbl_ext})
may be traced to a strong
departure of the modeled dust temperature along this direction
(D. Finkbeiner, priv. comm.), justifying avoidance of their use here.

For each
of the NICE, NICER, and RJCE analyses, extinction maps across the
T-Galaxy region were also developed. These did not show large gradients 
in extinction, nor particular holes or peaks at the T-Galaxy position. The
maps have higher uncertainties, however, as only a handful of stars contribute
information at each map location. Given the lack of new insight offered
by these maps, the region mean values listed above (and in Table~\ref{tbl_ext}) 
were adopted for all corrections.


\clearpage

\clearpage

\clearpage

\clearpage

\clearpage

\clearpage

\clearpage

\clearpage

\clearpage

\clearpage

\begin{deluxetable}{ccccccc}
\tabletypesize{\scriptsize}
\tablecaption{Polarization and Color Excess Properties by Region\label{tbl_p_vs_ehk}}
\tablewidth{0pt}
\tablehead{
\colhead{Region} & \colhead{N} & \colhead{$\sigma_{\rm P}$}
&\colhead{$\sigma_{\rm E(H-K)}$}&\colhead{$E(H - K)$}&\colhead{$<$P$>$}&
\colhead{$<E(H - K)>$}\\
&\colhead{(pixels)}&\colhead{(\%)} & \colhead{(mag)} & \colhead{(mag)} & 
\colhead{(\%)} &\colhead{(mag)}
}
\startdata
Core		&	4&	$<$~0.12&	$<$~0.02&	... 		&	0.83 (0.25)&	0.215 (0.017) \\
Bulge	&   26&	0.12 - 0.2&	$<$~0.03&	...		&	0.96 (0.21)&  0.192 (0.011) \\
Interarm &   82&   0.2 - 0.7&	$<$~0.10& $<$~0.17	& 	1.26 (0.14) & 0.110 (0.011)\\
Arms      &   61&      "       &    "	&	$>$~0.17 		& 	1.10 (0.15) & 0.229 (0.011) \\
Outer     &  123&  0.7 - 1.2 & $<$~0.20 &  ...			& 	1.08 (0.14) & 0.135 (0.015)\\
\hline
\enddata
\end{deluxetable}

\clearpage

\begin{deluxetable}{cccccl}
\tabletypesize{\scriptsize}
\tablecaption{Extinction Estimates toward the T-Galaxy\label{tbl_ext}}
\tablewidth{0pt}
\tablehead{
\colhead{Method}&\colhead{Quantity}&\colhead{Value}&\colhead{Equiv. A$_{\rm V}$}&
\colhead{$E(H - K)$}&\colhead{Ref.\tablenotemark{a}}\\
&&&\colhead{(mag)}&\colhead{(Mag)}
}
\startdata
IRAS/COBE & A$_{\rm H}$ & 0.676 mag& 4.127  & 0.260 & SF11 \\
{\it IRAS/COBE}& {\it A$_{\rm H}$ }&{\it  0.867 mag}& {\it 4.989}	& {\it 0.314} & {\it SFD98} \\
CO/$^{13}$CO & N$_{\rm H_2}$ & 2 - 2.5$ \times 10^{21}$ cm$^{-2}$ & 2 - 2.5 & 0.13 - 0.16 & GHN08 \\
NICE   & $A_{\rm V}$     & 1.80$\pm$0.17 mag  & 1.80 & 0.113$\pm$0.011 & \\
NICER & A$_{\rm K}$  & 0.10 - 0.15 mag        & 1.8 - 2.8 & 0.11 - 0.18 & LLA10 \\ 
NICER/3 & A$_{\rm V}$  & 2.18$\pm$0.08 mag  & 2.18 & 0.137$\pm$0.005 & \\
NICER/5 & A$_{\rm V}$  & 1.97$\pm$0.09 mag  & 1.97 & 0.124$\pm$0.006 & \\
RJCE      & A$_{\rm V}$  & 1.96$\pm$0.08 mag  & 1.85 & 0.123$\pm$0.005 &\\
\hline
Adopted &&&2.00$\pm$0.10 &0.125$\pm$0.009&
\enddata
\tablenotetext{a}{SF11: Schlafly \& Finkbeiner 2011 (from NED); SFD98: Schlegel, Finkbeiner \& Davis 1998 (from NED); GHN08: Goldsmith et al. (2008) - our visual interpolation; 
LCA10: Lombardi, Lada, \& Alves - our visual interpolation}
\end{deluxetable}

\clearpage

\end{document}